\documentclass{article}

\usepackage{xcolor}         
\usepackage{PRIMEarxiv}
\usepackage[utf8]{inputenc} 
\usepackage[T1]{fontenc}    
\usepackage{url}            
\usepackage{booktabs}       
\usepackage{amsfonts}       
\usepackage{amsmath}        
\usepackage{nicefrac}       
\usepackage{microtype}      
\usepackage{xcolor}         
\usepackage{algorithm}
\usepackage{algorithmic} 
\usepackage{lipsum}
\usepackage{natbib}
\usepackage{fancyhdr}       
\usepackage{graphicx}       
\graphicspath{{media/}}     
\usepackage{verbatim} 
\usepackage{amsfonts} 
\usepackage{hyperref} 
\hypersetup{
    colorlinks=true,
    urlcolor=blue
}
\usepackage{amssymb}

\title{Flow-TAG: Flow-based conditional latent transport for accurate spline approximation and~data~compression
}

\author{
  Roman Pavelkin, Luis A. Zavala-Mondragon, Fons van der Sommen \\
  Eindhoven University of Technology \\
  Eindhoven\\
  \texttt{\{r.pavelkin, l.a.zavala.mondragon, fvdsommen\}@tue.nl} \\
}

\begin{document}
\maketitle

\begin{abstract}

Robust curve fitting is essential in computer-aided design for transforming noisy, discrete data into accurate geometric models that ensure numerical stability across engineering workflows. B-spline models have become the industry standard for this task, offering a flexible and reliable framework characterized by local control and smooth shape representation. This paper presents flow-TAG--a data-driven framework based on a generative flow model with a 1D U-Net backbone capable of mapping the geometry of a curve to the optimal parametrization for cubic B-splines. By leveraging learned geometric patterns, flow-TAG exhibits superior parameterization performance, robustness to noise in the input data, and strong generalization capability to previously unseen 2D and 3D curves drawn from distinct data distributions. Flow-TAG yields fitted curves that achieve the lower root-mean-square error (55\% lower on average) and Hausdorff distance (52\% lower on average) relative to state-of-the-art data-driven methods. In addition, we investigate the practical applicability of our generative framework in the compression of ECG signals for wearable devices. The proposed compression setup provides a compression ratio of 13 with the signal distortion of $\approx5\%$, which is acceptable in the field.

\end{abstract}

\keywords{Curve fitting \and B-spline approximation optimization \and flow matching \and generative models \and latent transport \and ECG compression}

\section{Introduction}
\label{intro}


Robust curve fitting is crucial in Computer-Aided Design~(CAD) as it underpins the creation of geometric models essential for design and manufacturing~\citep{pottmann_industrial_2005}. Curves are typically derived from discrete data which can suffer from noise and irregular sampling. Effective curve fitting is necessary to accurately represent shapes, ensuring smoothness and numerical stability. However, it can be noted that poor fitting can introduce errors and undermine design intent. This reliability is vital for applications in surface modeling~\citep{farin_curves_2001}, tool path generation~(\cite{langeron_new_2004, lasemi_recent_2010}), and finite element analysis~\citep{hughes_isogeometric_2005}, making robust curve fitting methods essential for efficient engineering workflows.


Spline-based models have become a standard tool for curve approximation in CAD due to their flexibility, numerical stability, and strong theoretical foundation~\citep{farin_curves_2001}. Among them, B-splines are particularly prominent because they provide a compact representation with desirable properties such as local support, smoothness control, and affine invariance~\citep{prautzsch_bezier_2002}. A B-spline curve is defined by a set of control points, a polynomial degree, and a non-decreasing knot vector, which together determine the shape of the curve. The parametrization of the curve—i.e., the assignment of parameter values to data points—plays a crucial role in the quality of approximation, as it influences how the basis functions are evaluated and how geometric features are distributed along the curve. Common parametrization strategies include uniform, chord-length, and centripetal methods, each offering different trade-offs between simplicity and fidelity to the underlying geometry~\citep{piegl_nurbs_2012}. The combination of flexible parametrization and the inherent locality of B-spline basis functions makes spline-based approaches particularly well-suited for approximating complex shapes encountered in real-world design tasks.


In recent years, data-driven approaches have been increasingly explored to improve curve parametrization for B-spline fitting, addressing limitations of classical heuristic methods such as uniform or chord-length parametrization~(\cite{laube_learnt_2018, song_hybrid_2024}). Machine learning techniques, particularly neural networks, enable the estimation of parameter values directly from data by learning underlying geometric patterns and noise characteristics~\citep{lecun_deep_2015}. These methods aim to produce parametrizations that lead to improved approximation accuracy and robustness, especially for complex or highly irregular curves. A dedicated Section~\ref{rel_work_1} later in the manuscript provides a detailed discussion of the principal contributions on neural parametrization for B-spline fitting; therefore, we do not elaborate further on this topic at this point.


Within the scope of the stated problem, the aim of this study is to develop a data-driven framework capable of mapping the geometry of a curve to the optimal positions of spline knots by leveraging learned geometric patterns. In such a model, the knot vector serves as a latent representation of the curve being approximated.


We demonstrate how, using a generative flow-based model with a 1D U-Net backbone, it is possible to learn the geometry of 2D and 3D curves and map it into a knot vector. Ultimately, we introduce \textbf{flow-TAG}: \textbf{flow}-based \textbf{T}ask-\textbf{A}ware \textbf{G}enerative framework for conditional latent transport for accurate spline approximation. The framework combines several original design solutions, including: 1)~redefining the B-spline functional via a 1D U-Net deep learning model; 2)~training with the self-tuning multi-task geometry-informed loss function, which combines both the flow matching term and the curve fidelity term (task loss). As a result, flow-TAG exhibits superior parameterization performance (the resulting knot vectors produce the highest-quality spline approximation among all evaluated baseline methods), robustness to noise in the input data, and strong generalization capability to previously unseen curves drawn from distinct data distributions.


In addition to evaluating the intrinsic accuracy of the approximations of abstract curves, we also investigate the practical applicability of flow-TAG to real medical data that can be represented by splines. Specifically, we employ flow-TAG as the backbone to perform compression of ECG recordings~\citep{moody_impact_2001}. The results of our tests clearly demonstrate the potential of using flow-TAG in the compression of ECG signals for wearable devices.

The body of the paper consists of the following parts: Section~\ref{problem} includes the rigorous problem statement; Section~\ref{rel_work} outlines the baselines of the data-driven neural B-spline curve fitting and curve compression methods; Section~\ref{methods} provides the theoretical underpinning of the proposed framework, description of the dataset used in the study, and essential notes on the experiments; Section~\ref{res_disc} contains the experimental outcomes and their discussion; lastly, in Section~\ref{concl}, we analyze the impact of the study, discuss the limitations of flow-TAG, and potential future research interests in this regard. The paper is accompanied by Appendices that provide additional methodological and implementation details for interested readers.

\section{Problem statement}
\label{problem}

Let a curve be represented by a set of sampled points:
\begin{equation}
    \{(k_i, \mathbf{y}_i)\}_{i=1}^{N}, \quad \mathbf{y}_i \in \mathbb{R}^d,
    \label{eq:eq_1}
\end{equation}
where $k_i$ are ordered parameter values and $\mathbf{y}_i$ are the corresponding observations of a \textit{d}-dimensional signal. The goal is to construct a smooth function that accurately represents the underlying curve while maintaining a compact and flexible representation.

A widely used representation for such tasks is the B-spline curve~\citep{farin_curves_2001}. Given a non-decreasing knot vector
\begin{equation}
    \mathbf{K} = \{ k_{0}, k_{1}, \ldots, k_{m} \},
    \label{eq:eq_2}
\end{equation}
and a spline degree \textit{p}, the B-spline basis functions $B_{i,p}(k)$ are defined recursively using the Cox–de Boor formula~\citep{de_boor_practical_1978}:
\begin{equation}
    B_{i,0}(k) = 
    \begin{cases}
        1, & k_i \le k < k_{i+1}, \\
        0, & \text{otherwise},
    \end{cases}
    \label{eq:eq_3}
\end{equation}
\begin{equation}
    B_{i,p}(k) = \frac{k - k_i}{k_{i+p} - k_i} B_{i,p-1}(k) + \frac{k_{i+p+1} - k}{k_{i+p+1} - k_{i+1}} B_{i+1,p-1}(k).
    \label{eq:eq_4}
\end{equation}
A B-spline curve of degree \textit{p} is then expressed as
\begin{equation}
    \mathbf{S}(k) = \sum_{i=0}^{n} c_i B_{i,p}(k),
    \label{eq:eq_5}
\end{equation}
where $\mathbf{c}_i \in \mathbb{R}^d$ are the control points, and the number of basis functions is determined by the knot vector length via $m = n + p + 1$.

B-splines were originally constructed as convolutions of certain probability distributions and later used for statistical data smoothing~\citep{farin_curves_2001}. This is remarkable because B-spline basis functions exhibit a structural resemblance to point spread functions (PSF) in optics and kernels in kernel density estimation (KDE). Just as a PSF describes the spatial distribution of a point source’s energy due to system blur, or a KDE kernel distributes the probabilistic "weight" of a discrete data point, the B-spline basis function $B_{i,p}(k)$ serves as a localized weighting operator that distributes the influence of a control point across a specific interval. Mathematically, this relationship is rooted in the fact that a B-spline of order $p+1$ ($p$ is degree) can be derived from the $p+1$-fold convolution of a rectangular pulse (the boxcar function), a process that mirrors the repeated filtering operations used to model optical blur or to derive higher-order smoothing kernels~(\cite{de_boor_practical_1978, unser_splines_2002}). Consequently, B-spline interpolation can be interpreted as a generalized filtering process where the continuous signal is reconstructed by "smearing" discrete coefficients through a basis that satisfies the partition of unity, ensuring both local support and global smoothness, making B-spline curves, Eq.~\ref{eq:eq_5}, a powerful tool to model complex shapes.

The curve modeling can be carried out as an approximation or an interpolation process. In the presented study, both frameworks were utilized. In the spline interpolation problem, the constructed curve is required to pass exactly through the given data points:
\begin{equation}
    \mathbf{S}(k_i) = \mathbf{y}_i, \quad i=1, \ldots, N.
    \label{eq:eq_6}
\end{equation}
Substituting the B-spline representation yields a linear system
\begin{equation}
    \sum_{j=0}^{n} c_j B_{j,p}(k) = \mathbf{y}_i;
    \label{eq:eq_7}
\end{equation}
or in matrix form
\begin{equation}
    \mathbf{B} \mathbf{C}=\mathbf{Y},
    \label{eq:eq_8}
\end{equation}
where $\mathbf{B}_{ij}=B_{j,p}(k)$ is the basis matrix, $\mathbf{C}$ contains the unknown control points, $\mathbf{Y}$ contains the observed samples. If the number of basis functions equals the number of samples, $n+1=N$, the system~(\ref{eq:eq_8}) can be solved directly. Interpolating splines provide exact reconstruction of the observed data but may become sensitive to noise or irregular sampling.

In many practical applications the number of control points is intentionally chosen to be smaller than the number of samples, $n+1<N$. In this case, the spline does not interpolate the data exactly but approximates it by minimizing a fitting error. A common formulation is the least-squares problem:
\begin{equation}
    \min_{\mathbf{C}} \sum_{i=1}^{N} \left\lVert y_i - \sum_{j=0}^{n} c_j B_{j,p}(k_i) \right\rVert^{2};
    \label{eq:eq_9}
\end{equation}
in matrix form:
\begin{equation}
    \min_{C} \lVert Y - BC \rVert^{2};
    \label{eq:eq_10}
\end{equation}
the solution is given by
\begin{equation}
    (\mathbf{B}^\top \mathbf{B}) \mathbf{C} = \mathbf{B}^\top \mathbf{Y}.
    \label{eq:eq_11}
\end{equation}
Approximation B-splines provide a compact representation of the curve and naturally smooth noisy observations.

The knot vector $\mathbf{K}$ determines the support and shape of the B-spline basis functions and therefore strongly influences the quality of the curve representation. For a fixed spline degree \textit{p} and number of control points, the choice of knot locations defines the approximation space
\begin{equation}
    \mathcal{S}_{p,\mathcal{K}} = \operatorname{span}\{B_{0,p}, B_{1,p}, \ldots, B_{n,p}\}.
    \label{eq:eq_12}
\end{equation}
However, the knots are in general not given, which poses the problem of finding an optimal knot placement scheme for a given curve. Uniform knot placement is often used due to its simplicity, but it may fail to capture local variations of the curve, particularly in regions with high curvature or rapid changes. Conversely, adaptive placement of knots allows more flexible modeling by allocating more basis functions in complex regions and fewer in smooth segments.

Formally, the optimal knot placement problem can be formulated as
\begin{equation}
    \min_{\mathbf{C}} \sum_{i=1}^{N} \left\lVert y_i - \sum_{j=0}^{n} c_j B_{j,p}(k_i; \mathbf{K}) \right\rVert^{2},
    \label{eq:eq_13}
\end{equation}
subject to the monotonicity constraint on the knot vector
\begin{equation}
    k_0 \leq k_1 \leq \cdots \leq k_m.
    \label{eq:eq_14}
\end{equation}
Moreover, in the present study, we imposed natural boundary conditions at the extremities of the knot vector, which effectively means the second derivatives at the ends are set to zero.

The problem given by Eq.~(\ref{eq:eq_13}) is nonlinear and combinatorial because the basis functions themselves depend on the knot locations. Consequently, determining an optimal knot configuration is a challenging task and often relies on heuristic strategies, iterative refinement, or data-driven approaches. A brief survey of approaches for adaptive knot placement for B-spline curve approximation can be found in Section~\ref{rel_work_1}.

Efficient spline modeling therefore requires addressing two coupled tasks: estimating the spline coefficients for a given knot configuration and determining an appropriate placement of knots that yields an accurate yet compact representation of the underlying curve.

\section{Related work}
\label{rel_work}

In this section, we provide an overview of several notable previously published works related to the optimization of spline curve approximation using machine learning, as well as studies on curve compression algorithms.

\subsection{Learning-based methods for spline curve approximation}
\label{rel_work_1}

Addressing optimal parameter and knot estimation in B-spline curve fitting, Laube~et~al.~\cite{laube_deep_2018} proposed a data-driven framework called PARNET, utilizing deep neural networks instead of traditional heuristics. PARNET features two architectures: a Point Parametrization Network (PPN) for assigning parametric values, and a Knot Selection Network (KSN) for predicting knot vector distribution. This approach allows simultaneous prediction of parameters and knots based on point set geometry, incorporating a least-squares control point solution into the network’s loss function for reduced approximation error. Results show PARNET achieves tighter fits with fewer control points and lower computational overhead during inference. However, challenges include scalability and generalization, as training on synthetic data limits robustness against real-world noise distributions.

Unlike PARNET, which used fixed-size global inputs, Scholz~et~al.~\cite{scholz_parameterization_2021} employ a Residual Neural Network (ResNet) to process a small, fixed window of points necessary for unique polynomial approximations. This approach allows the model to generalize to arbitrary-length point sequences and minimize approximation error, effectively capturing non-linear relationships and outperforming traditional methods. However, it faces challenges like vulnerability to high-frequency noise and requires retraining for different polynomial degrees, relying on pre-ordered inputs. Increasing the number of layers and knots may be required for approximating complex curves in practical applications.

Wen~et~al.~\cite{wen_deep_2024} in their paper introduced a DNN-Solver for B-spline approximation. In their framework, a Deep Neural Network (DNN) was used as a means of mapping knot positioning from a random allocation to a suboptimal space given the curve data points. The input can be any initial fixed knots, and the output provides the desirable knots. This constitutes a relatively straightforward, predominantly optimization-based rather than learning-based approach. However, its generalizability property necessitates a computationally expensive iterative optimization procedure, which precludes the application of DNN-Solver to real-time curve-fitting scenarios.

With the emergence and widespread adoption of transformer-based deep neural network architectures in recent years, which have achieved state-of-the-art performance across a broad range of tasks, their application to the B-spline curve approximation problem became a natural development. An example of such an approach is presented in the work of~Saillot~et~al.~\cite{saillot_b-spline_2024}. Trained with supervised learning on randomly generated B-spline curves, their method is designed to directly output the knot vector. Another generative transformer-based approach called SplineGen was introduced in the paper of~Zou~et~al.~\cite{zou_splinegen_2025}. It addresses challenges both in parameterization and knot placement in B-spline approximations. Its key features include independence from input data point size, permutation invariance, aligned knots and parameters, and robustness, leading to highly accurate curve fittings. The limitations of SplineGen include failing in scenarios involving point disorder and lack of explicit mechanisms for constraining approximation errors or ensuring a minimal number of knots. 

\subsection{Curve compression techniques}
\label{rel_work_2}

There are diverse techniques that can be used to achieve compression of curves. Among such examples we can find spline curve fitting techniques.  For example, previous work~\citep{pavelkin_spline-based_2025} proposed an heuristic algorithm for compressing the 3D shape of medical guidewires based on the natural cubic spline approximation with an optimized knot placement procedure given a geometry prior. This approach demonstrated robustness in the efficient representation of complex, high-resolution medical shape measurements. Nevertheless, due to its reliance on an optimization-based procedure, the majority of the compression time is devoted to determining an optimal knot placement, which scales poorly with problem size and consequently induces a trade-off between real-time performance and achievable~compression~ratio.

Another example of curve fitting compression has been used for accurate B-spline fitting and compression of ECG~\citep{mohebbian_ecg_2023}. The referred method employs stochastic ant colony optimization to find optimal knot locations along the ECG segments. Analogously to the previously discussed approach, this method depends on a computationally intensive optimization procedure, which constrains its suitability for real-time applications. Furthermore, the method may exhibit limited robustness in the presence of noisy recordings~\citep{mohebbian_ecg_2023}.

Transform-based compression techniques filter insignificant components out by enforcing sparsity in the transform domain. Analogously to image compression, Bilgin~et~al.~\cite{bilgin_compression_2003} demonstrated how JPEG2000 codec can be applied to effectively compress and denoise ECG recordings. However, the method introduces significant distortion to the signal after decompression. Another strong baseline based on the combination of the coalition of empirical mode decomposition and Tunable-Q Wavelet Transform~(TQWT) was developed by~Sharma~et~al.~\cite{sharma_ecg_2024}. The method benefits from packing the maximum energy with fewer wavelet coefficients, which have a significant contribution to the original signal, and dynamic thresholding and dead-zone quantization to discard the insignificant wavelet coefficients. Subsequently, a run-length encoding lossless compression scheme was employed to encode the retained wavelet coefficients. The strengths in impressive compression rate with moderate distortions attributed to this algorithm come with the following limitations: EMD-based approaches suffer from high computational complexity and the mode mixing problem~(\cite{chang_arrhythmia_2010, xu_ecg_2017}), while TQWT requires careful manual parameter tuning that reduces generalizability~(\cite{singhai_ecg_2023, pal_electrocardiogram_2023}). More broadly, these pipelines are typically evaluated on a single, limited database, lack end-to-end optimization, and may discard clinically significant signal features through aggressive thresholding.

Yildirim~et~al.~\cite{yildirim_efficient_2018} have presented a pioneering work that uses Convolutional AutoEncoders~(CAE) for ECG compression. The encoder in their model consists of 14 layers, and the compression is achieved via dimensionality reduction at the bottleneck as compared to the original signal space. Their method achieves better average compression ratio relative to the traditional methods; however, the ratio is statically predefined and tightly coupled to the model architecture. Regardless of the impressive compression ratio, the method of~Yildirim~et~al.~\cite{yildirim_efficient_2018} reconstructs the signal with noticeable distortions higher than those of traditional methods and varies from signal to signal. The recent development of the concept of CAE for ECG compression was presented in the study of~Bekiryazici~et~al.~\cite{bekiryazici_novel_2025}, where they reported a multichannel CAE-based model. The proposed approach encodes the ECG signal into a four-channel low-dimensional space with increasing levels of sparsity constraints imposed on each channel. Different quantization levels are then applied to each channel with subsequent entropy encoding. Their method demonstrated the capability to efficiently compress ECG signals while maintaining reconstruction accuracy.

\section{Methods}
\label{methods}

This section exposes the methodology underlying the numerical experiments, including the architecture and configuration of the neural networks, the characteristics and preparation of the datasets, and all additional implementation details required to ensure a comprehensive understanding of the study and to facilitate full reproducibility of the reported results.

\subsection{Deep neural network to substitute the spline approximation}
\label{methods_1}

Given the problem~(\ref{eq:eq_9}) and the smoothness penalty
\begin{equation}
    \lambda \int \lVert \mathbf{y}''(t) \rVert^{2} \, dt
    \label{eq:eq_15}
\end{equation}
imposed by the natural boundary conditions applied to the endpoints $y_0$ and $y_n$, the spline approximation can be regarded as a smoothing low-dimensional representation ($\lambda>0$ is a smoothing parameter). The penalty given by Eq.~(\ref{eq:eq_15}) actually reflects the minimum property of cubic splines with natural boundary conditions, which have intuitive physical interpretation: the spline model mimics the behavior of a thin, flexible elastic beam, minimizing bending energy~\citep{farin_curves_2001}. Therefore, the objective of such a smoothing spline functional becomes:
\begin{equation}
    \min_{f} \sum_{i} (y_i - f(t_i))^2 + \lambda \int (f''(t))^2 \, dt,
    \label{eq:eq_16}
\end{equation}
which is also known as a \textit{variational formulation of spline fitting}~\citep{prenter_splines_2008}.

We define the functional $f$ from Eq.~(\ref{eq:eq_16}) by a neural network $f_\theta$ parametrized by $\theta$. Such a network learns a mapping
\begin{equation}
    f_{\theta} : y_{\text{masked}} \rightarrow y_{\text{reconstructed}},
    \label{eq:eq_17}
\end{equation}
where
\begin{equation}
    y_{\text{masked}} = M \odot y; M \in \{0,1\}^N,
    \label{eq:eq_18}
\end{equation}

Therefore, a network $f_\theta$ trained to minimize loss, which includes the objective given by Eq.~(\ref{eq:eq_16}), is encouraged to produce curves that both match the observed samples and have small curvature matching the curvature of the observed samples. In other words, the network is learning a function that approximates the solution of a smoothing spline problem, and therefore can be interpreted as a learned nonlinear and differentiable generalization of spline smoothing with curvature regularization. Having established the mathematical rationale for moving from the B-spline approximation to a more general neural network model $f_\theta$, we will now define the principal architecture of such a network. 

\begin{figure}[tb]
    \centering
    \includegraphics[width=0.8\linewidth]{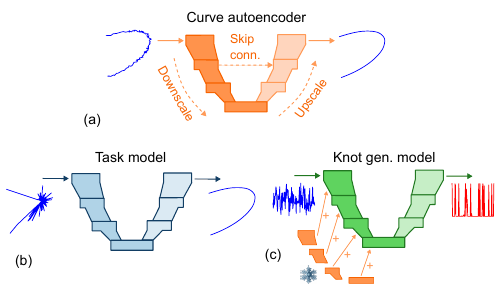}
    \caption{Diagrams of the 1D U-Net-based neural networks used in the study: (a) autoencoder $\mathcal{D}$ for curve denoising and multiscale feature extraction; (b) task model $f_\theta$ mimicking B-spline approximation; (c) flow-based soft knot-mask generative model $v_\phi$.}
    \label{fig_1}
\end{figure}

A natural architectural choice for this problem is a one-dimensional U-Net–based autoencoder, since several aspects of its structure align well with the theory of spline approximation and smoothing. U-Nets are particularly suitable for inpainting and denoising tasks because they are explicitly designed for context-aware reconstruction of signals or images~(\cite{ronneberger_u-net_2015, williams_unified_2023}), which makes them a good fit for reconstructing curves from sparse samples (as formulated by Eq.~(\ref{eq:eq_17})). A central component of U-Net is the system of skip connections between encoder and decoder layers (see Figure~\ref{fig_1}~(a)), which enables the decoder to merge local high-resolution details (fine structure) with global contextual information (shape constraints). This is especially beneficial for curve reconstruction: the network can preserve the exact values of observed samples while filling in unobserved regions using global context. In this way, skip connections help maintain local geometric constraints, analogous to how spline bases provide local support around knots. In addition, U-Net handles noisy input data effectively, allowing the mapping function \(f_\theta\) to adapt to the data distribution and remain robust to noise. For these reasons, U-Net serves as the primary architecture in this work; its integration into the flow-based generative framework is described in the next subsection.

\subsection{Task-aware flow-based generative modeling}
\label{methods_2}

This section outlines the essential aspects of the proposed \textbf{flow}-based \textbf{T}ask-\textbf{A}ware \textbf{G}enerative~(flow-TAG) framework for conditional latent transport (knot vector). We first describe the origin of this framework and its applications. Next, we present the adaptation for curve approximation and discuss the multi-task geometry-aware loss.

\subsubsection{Conventional Flow Matching and its coupling with a downstream task}
\label{methods_2_1}

\begin{figure}[tb]
    \centering
    \includegraphics[width=0.9\linewidth]{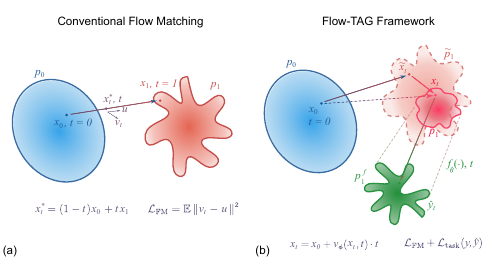}
    \caption{Workflows of the (a) conventional flow matching set-up and (b) proposed flow-based framework for generating latent $x_t$ for a task $f_\theta$.}
    \label{fig_2}
\end{figure}

As a distribution-matching approach, Flow Matching (FM) learns a time-dependent vector field $v_\phi(x, t)$ within a neural ODE
\begin{equation}
    \frac{dx}{dt} = v_{\phi}(x_t, t), \quad x(0) = x_0,
    \label{eq:eq_19}
\end{equation}
\begin{equation}
    x^{*}_{t} = x_0 + v_\phi(x_t, t) \cdot t,
    \label{eq:eq_20}
\end{equation}
that deterministically maps samples $x_0$ from the standard Gaussian distribution~(noise) $p_0$ to the target distribution $p_1$ (see Figure~\ref{fig_2}~(a)) (\cite{lai_principles_2025, lipman_flow_2022}). The training objective of a FM model $v_\phi$ is to minimize the $L^2$ regression loss between the learned velocity field $v_{\phi}(x_t, t)$ and the ground-truth velocity $u$, which specifies the true trajectory of each point as it evolves from $p_0$ to $p_1$:
\begin{equation}
    \mathcal{L}_{\text{FM}} = \mathbb{E}\left\| v_{\phi}(x_t, t) - u \right\|_2^2,
    \label{eq:eq_21}
\end{equation}
where $u$ is typically expressed through the linear coupling assumption between $p_0$ and $p_1$. Thus, at a time step $t$, the path $x^{*}_t$ is defined as
\begin{equation}
    x^{*}_t = (1 - t)\,x_0 + t\,x_1, \quad t \in [0,1],
    \label{eq:eq_22}
\end{equation}
where $x_0 \in p_0$ and $x_1 \in p_1$. As the formulation~Eq.~(\ref{eq:eq_22}) defines a straight line, each pair $(x_0, x_1)$ has a constant ground-truth velocity $u = \frac{d}{dt} x_t = x_1 - x_0$ along the path $x_t$. 

Let us consider a setting in which the ground-truth distribution \(p_1\) is not directly accessible and only an empirical or approximated estimate \(\tilde{p}_1\) can be obtained. At the same time, we posit the hypothesis that the true distribution either coincides with the estimated one or is contained within the family of distributions represented by \(\tilde{p}_1\):
\begin{equation}
    \text{supp}(p_1) \subseteq \text{supp}(\tilde{p}_1),
    \label{eq:eq_23}
\end{equation}
or overlaps with it:
\begin{equation}
    \text{supp}(p_1) \cap \text{supp}(\tilde{p}_1) \neq \emptyset.
    \label{eq:eq_24}
\end{equation}

Given that the samples from both distributions are latent representations for some function (in our case, it is the spline approximant $f_\theta$, Section~\ref{methods_1}, and the target distribution $p_1$ would be then the space of possible knot positions), $p_1$ can be estimated indirectly via the predictions of the model $f_{\theta}(x_{t}) \to \hat{y}$. Therefore, the flow prediction~(Eq.\ref{eq:eq_20}) becomes a latent representation of the task $f_\theta$. Once both the flow $v_\phi$ and the task $f_\theta$ networks are differentiable, learning the transformation dynamics $x_0 \to x_1$ becomes an end-to-end optimization problem driven by the downstream loss $\mathcal{L}(y, \hat{y})$:
\begin{equation}
    \frac{\partial \mathcal{L}}{\partial \phi} = \frac{\partial \mathcal{L}}{\partial \hat{y}} \cdot \frac{\partial \hat{y}}{\partial x_t} \cdot \frac{\partial x_t}{\partial \phi}.
    \label{eq:eq_25}
\end{equation}
Figure~\ref{fig_2}~(b) shows a diagram of this generative framework. The time-dependent weighting in Eq.~(\ref{eq:eq_20}) effectively functions as an implicit regularizer: at early times $(t \to 0)$, the flow emphasizes coarse, broadly data- and task-independent structure, whereas at later times $(t \to 1)$ it encodes task-specific refinements. The B-spline approximant $f_\theta$ fixes the terminal state of the flow within a meaningful objective space, thereby acting~as~a~prior.

\subsubsection{Multi-task loss with uncertainty-based weighting}
\label{methods_2_2}

The training objective of flow-TAG was to minimize the following loss function: 
\begin{equation}
    \mathcal{L}_{\text{total}}(\phi,\sigma_{1},\sigma_{2},\sigma_{3})=\underbrace{\frac{1}{2\sigma_1^{2}}\mathcal{L}_{\text{FM}}(\phi)}_{\text{FM Loss}} + \underbrace{\frac{1}{2\sigma_2^{2}}\mathcal{L}_{\text{MSE}} + \frac{1}{2\sigma_3^{2}}\mathcal{L}_{\text{Smooth}}}_{\text{Geometry-informed Reconstruction Loss}} + \log\sigma_{1}\sigma_{2}\sigma_{3},
    \label{eq:eq_26}
\end{equation}
which consists of three separate loss terms and optimizable weights $\sigma$. The idea of optimizing the coefficients of the terms that make up the loss function was proposed in the work of~Kendall~et~al.~\cite{kendall_multi-task_2018}. This approach makes it possible to balance the model training process on complex composite tasks such as ours by learning the relative weight of each task adaptively. As $\sigma$ increases, the weight of the corresponding objective $\mathcal{L}$ decreases, and vice versa. Weights are discouraged from increasing too much by the last logarithmic term in Eq.~(\ref{eq:eq_26}) in order to avoid the risk of ignoring a certain objective. Therefore, $\mathcal{L}_{\text{total}}$ becomes effectively “self-tuning”. We now analyze each individual objective appearing in Eq.~(\ref{eq:eq_26}).

As mentioned in Section~\ref{methods_2_1}, Eq.~(\ref{eq:eq_21}), the first term of the total loss is responsible for the overall consistency of the generated latent representations with the estimated ground truth distribution (flow $p_0\rightarrow\tilde{p}_1$, Figure~\ref{fig_2}~(b)). To evaluate this term, it is both necessary and sufficient to possess an approximate characterization of the latent distribution. Let this pseudo–ground truth be represented by the function $\mathcal{F}$. In our work, we instantiate $\mathcal{F}$ using the knot–placement algorithm introduced by~Yeh~et~al.~\cite{yeh_fast_2020}. The authors proposed a heuristic procedure, which can be concisely summarized as follows:
\begin{enumerate}
    \item Calculate the derivatives of the input data.
    \item Calculate a feature curve for the input data.
    \item Determine the knot vector using the feature curve.
\end{enumerate}
However, the algorithm of Yeh~et~al.~\cite{yeh_fast_2020} only works in the absence of noise in the data. In the opposite case, its performance becomes unstable. This limitation, however, does not pose an issue in our setting, because the function~$\mathcal{F}$ is employed exclusively during the training phase of flow-TAG, during which a denoising U-Net model is concurrently utilized (see Figure~\ref{fig_1}~(a)).

Fine-tuning of the learned latent distribution is carried out by means of the task loss function - Geometry-informed reconstruction loss $\mathcal{L}(y, \hat{y})=\mathcal{L}_{\text{MSE}}+\mathcal{L}_{\text{Smooth}}$. The first term minimizes the average reconstruction error
\begin{equation}
    \mathcal{L}_{\text{MSE}}=\mathbb{E} \left\| \hat{y} - y \right\|_2^2;
    \label{eq:eq_27}
\end{equation}
whereas the second term penalizes the $L^1$ error of the second spatial derivative of the reconstructed curve $\hat{y}$ to match that of the original curve $y$
\begin{equation}
    \mathcal{L}_{\text{Smooth}}=\mathbb{E} \left\lVert \nabla^2 \hat{y} - \nabla^2 y \right\rVert_1.
    \label{eq:eq_28}
\end{equation}

In this formulation of the task loss, we enforce equivalence between the learning objective and objective~(\ref{eq:eq_16}). This alignment justifies the assertion that the soft masks produced by flow-TAG reside in a latent space that is optimal for B-spline approximation, which follows as well from the ablation study given in~\ref{app_2}.

\subsubsection{Differentiable sigmoid gate}
\label{methods_2_3}

The generative flow model \(v_\phi\) outputs a velocity field that transports the noise-distributed latent variables to the corresponding optimal latent distribution. However, the ground truth subsampling masks belong to binary space, Eq.~(\ref{eq:eq_18}), which poses the fundamental limitation on backpropagation through step-like functions while training the model. To circumvent the non-differential nature of discrete sampling, the latent output $x_t$ from $v_\phi$ undergoes the following sigmoid transformation:
\begin{equation}
    g(x_t) = \mathcal{S} \!\left( s \cdot (x_t - \tau) \right), \quad s=20t^2
    \label{eq:eq_29}
\end{equation}
where $\mathcal{S}(\cdot)$ is the sigmoid function, $\tau = \mathrm{Quantile}_p(x_t)$ is the percentile threshold for soft-selection of the top $M$ entities in $x_t$, $s$ is the heuristically defined time-dependent non-linear steepness function, $t$ is the time step. 

The output of Eq.~(\ref{eq:eq_29}) $g(x_t) \in (0,1)$ represents a soft, differentiable mask for the task: elements of $x_t$ above the chosen percentile $M$ receive values close to $1$, while elements below are smoothly suppressed toward $0$ (see the red output soft mask in Figure~\ref{fig_1}~(c)). The steepness function enhances the sigmoid transformation as the time step $t$ approaches 1 (hard thresholding), whereas times below 0.8 normally result in soft thresholding. The gated output $g(x_t)$ is then passed to $f_\theta$. At the inference, the top-M selection is applied directly to the generated latent $x_1$. The differentiable given in Eq.~(\ref{eq:eq_29}) is analogous to the Differentiable Top-K Operator from the paper of~\cite{zhu_differentiable_2025}.

\subsubsection{Flow-TAG overview}
\label{methods_2_4}

The overall framework comprises three 1D U-Net-based networks that share an identical architecture, as illustrated in Figure~\ref{fig_1}. The input signal (a 2D or 3D curve) is fed to the denoising autoencoder $\mathcal{D}$, which extracts the multiscale noise-robust features $h$ (4 levels of depth + bottleneck). These features are then passed to guide the flow model $v_\phi$, which generates knot placement logits. The task model $f_\theta$ is used only at the training stage of the flow model. A detailed description of the model architectures and training process is provided in~\ref{app_1}. The inference workflow of flow-TAG is summarized in Algorithm~\ref{alg:flow-tag_inference}.

\begin{algorithm}[tb]
\caption{Conditional latent transport for accurate
spline approximation with flow-TAG}
\label{alg:flow-tag_inference}
\begin{algorithmic}[1]
\REQUIRE Original curve $y \in \mathbb{R}^{N \times d}$, denoising autoencoder $\mathcal{D}$, flow model $v_\phi$, ODE time grid $\{t_i\}_{i=0}^{T-1}$, $T=\Delta t^{-1}$--time step, number of knots $M$, 
\STATE $h \gets \mathcal{D}(y)$ \quad \textcolor{gray}{// get conditioning features}
\STATE Sample $x_0 \sim \mathcal{N}(0, I)^{N \times 1}$
\STATE Get soft mask logits: 
$\hat{x} = x_0+\int_{0}^{1} v_\phi(x_t, t, h) \, \text{d}t$ \quad \textcolor{gray}{// solve conditional flow ODE}
\STATE $b \gets \text{Top-}K(\hat{x})$ \quad \textcolor{gray}{// keep $K=M$ largest elements, set others to 0}
\STATE $\hat{\mathbf{K}} \gets \text{supp}(\mathbf{b})$ \quad \textcolor{gray}{// get knot indices}
\STATE Fit cubic B-spline:
$S = \sum_{i=0}^{n} c_i B_{i,3}(\hat{k})$
\STATE Reconstruct curve: $\hat{y} \gets S_b$
\RETURN Reconstructed curves $\hat{y}$, selected knots $\hat{\mathbf{K}}$
\end{algorithmic}
\end{algorithm}

\subsection{Datasets}
\label{methods_3}

In this work, three datasets were used. The training, validation, and test sets used in this article were carried out on the public SplineGen dataset by~~Zou~et~al.~\cite{zou_splinegen_2025}. In addition, the assessment of flow-TAG’s performance in clinically relevant settings was conducted on the public electrocardiogram (ECG) MIT-BIH Arrhythmia dataset~(\cite{moody_impact_2001}), which is a popular benchmark in the ECG data compression community.

\begin{figure}[tb]
    \centering
    \includegraphics[width=0.8\linewidth]{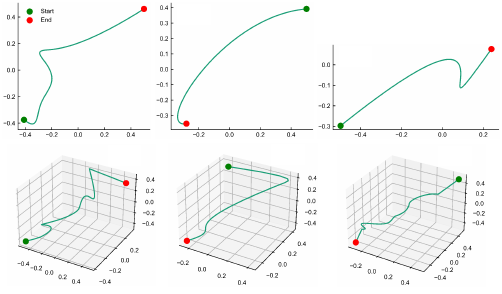}
    \caption{Exemplary 2D and 3D curves from the SplineGen dataset.}
    \label{fig_3}
\end{figure}

\paragraph{SpineGen dataset} This set consists of more than 1M non-self-intersecting B-spline-generated 2D and 3D curves. All curves were scaled and rotated to keep the pair of control points with the largest distance in the diagonal of the space ($[0, 1]^d$, \textit{d} is the dimension). The curves are flipped to avoid memorizing a positional bias. As knots and control points are provided within this dataset, all curves were resampled to be of the same length of 100 points. For our experiments, the dataset is split into train, validation, and test subsets. Examples of 2D and 3D curves from the SplineGen dataset are displayed in Figure~\ref{fig_3}.

\paragraph{MIT-BIH Arrhythmia Database} The database has been curated by the Massachusetts Institute of Technology and Beth Israel Hospital~(\cite{moody_impact_2001}). This dataset serves as a foundational benchmark for the development and validation of automated arrhythmia detection systems and compression algorithms for wearable devices. The database comprises 48 half-hour excerpts of two-channel (leads) ambulatory ECG recordings, obtained from 47 subjects studied between 1975 and 1979. The recordings have been digitized at a sampling rate of 360~Hz per channel with 11-bit resolution over~a~10~mV~range.

\begin{figure}[tb]
    \centering
    \includegraphics[width=1.0\linewidth]{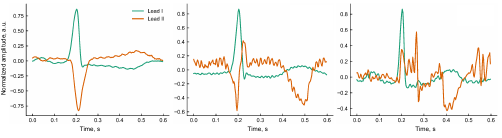}
    \caption{Exemplary 2-lead beats extracted from MIT-BIH Arrhythmia Database.}
    \label{fig_5}
\end{figure}

The acquired ECG signal comprises multiple noise and artifact components, including baseline drift, power-line interference, and high-frequency noise, all of which degrade signal quality and consequently impair the effectiveness of compression algorithms. To mitigate their impact, the ECG signals underwent the following preprocessing steps prior to being input into flow-TAG. As important clinical information in ECG signals generally lies in the frequency range of $1-150$~Hz, each ECG record was first passed through a 4th-order Butterworth filter with a bandwidth of $1-150$~Hz. Next, a band-stop notch filter with a center frequency of 50~Hz was applied to eliminate the noise originating from the power line. After the filtering, recordings were segmented into patches of 100 points followed by the mean-centered min–max normalization. Figure~\ref{fig_5} shows 3 examples of filtered and normalized ECG signal frames.
After preprocessing, ECG signal can be regarded as 2D curves and, therefore, processed by flow-TAG and approximated by~the~spline-fitting~model.

\subsection{Metrics}
\label{methods_4}

Given the input curves $y$ with $N$ points and their B-spline approximation $\hat{y}$ made with the flow-generated knots, we evaluated the performance of flow-TAG with the following most commonly used metrics: Root Mean-Squared Error~(RMSE)~(Eq.~(\ref{eq:eq_30})) and Hausdorff distance~(Eq.~(\ref{eq:eq_31})).
\begin{equation}
    \text{RMSE} = \sqrt{\frac{1}{N} \sum_{i=1}^{N} \left\| y_i - \hat{y}_i \right\|_2^2},
    \label{eq:eq_30}
\end{equation}
\begin{equation}
    \text{Haus. dist.} = \max \left\{ \max_{y \in Y} \min_{\hat{y} \in \hat{Y}} \lVert y - \hat{y} \rVert_2,\; \max_{y \in Y} \min_{\hat{y} \in \hat{Y}} \lVert \hat{y} - y \rVert_2 \right\}.
    \label{eq:eq_31}
\end{equation}
RMSE estimates the overall quality of the fit, and Hausdorff distance measures the dissimilarity between two curves by determining the maximum distance from any point in one curve to the nearest point in the other, and vice versa.

Next, in Section~\ref{res_disc_3}, when analyzing the quality of ECG signal approximation for compression, we evaluated the Compression Ratio~(CR) and the metrics that are more specific to the medical community of ECG signal processing~\citep{zigel_weighted_2002}. These metrics include Normalized Percent Root Difference (PRDN) and Quality Score (QS). Their~definitions~are:

\begin{equation}
    \text{CR} = \frac{\text{original file size}} {\text{compressed file size}},
    \label{eq:eq_32}
\end{equation}
\begin{equation}
    \text{PRDN} = \frac{\sqrt{\sum_{i=1}^{N} \left(y_i - \hat{y}_i\right)^2}}{\sqrt{\sum_{i=1}^{N} \left(y_i - \bar{y}\right)^2}} \times 100,
    \label{eq:eq_33}
\end{equation}
\begin{equation}
    \text{QS} = \frac{\text{CR}}{\text{PRDN}}.
    \label{eq:eq_34}
\end{equation}

PRDN is a quantitative metric that characterizes the deviation of a compressed signal from its original counterpart in percentage terms and is employed to assess the magnitude of reconstruction error introduced by the compression process. PRDN below a defined threshold is crucial for assessing whether a signal is suitable for clinical interpretation. Values between 0\% and 2\% reflect excellent compression. Prior work has shown that PRDN values under 9\% preserve all clinically relevant features with adequate signal quality~\citep{zigel_weighted_2002, bekiryazici_novel_2025}. In contrast, PRDN values above 9\% indicate that the reconstructed signal may lose critical information and thus be unsuitable for diagnostic use. QS evaluates the performance of ECG signal compression algorithms by jointly quantifying the achieved compression ratio and the magnitude of post-compression signal distortion. This metric emphasizes that methods attaining a high degree of data reduction while maintaining a low reconstruction error are regarded as exhibiting superior compression performance.

\section{Results and discussions}
\label{res_disc}

In this section, we present and analyze the results of numerical experiments: flow-TAG training dynamics, the quality of curve approximation with generated knots, and the performance of flow-TAG as compression algorithm for ECG.

\subsection{Training details}
\label{res_disc_1}

\begin{figure}[tb]
    \centering
    \includegraphics[width=0.5\linewidth]{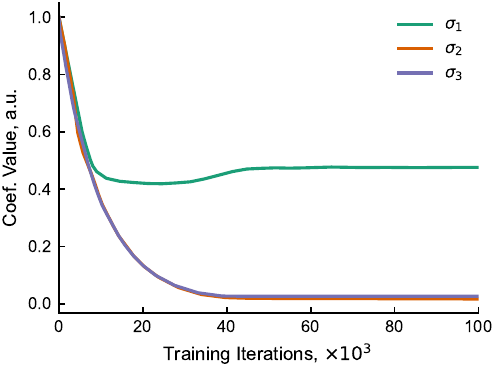}
    \caption{Training evolution of the weights $\sigma_{1,2,3}$ of the flow-TAG training objective.}
    \label{fig_6}
\end{figure}

We now examine the evolution of the weighting coefficients $\sigma_{1,2,3}$ associated with the terms of the total loss function in Eq.~(\ref{eq:eq_26}), as illustrated in Figure~\ref{fig_6}. At the initial stage of training, all three weights decrease at approximately the same rate. After roughly 10,000 iterations, however, the update of $\sigma_{1}$ slows down and, following a short adaptation phase, it stabilizes at a value of approximately 0.48. In contrast, $\sigma_{2}$ and $\sigma_{3}$ continue to decrease up to about 40,000 iterations, reaching values of 0.02 and 0.03, respectively.

Concurrently, the coefficients multiplying the $\mathcal{L}_{\text{MSE}}$ and $\mathcal{L}_{\text{Smooth}}$ terms increase as an inverse quadratic function of $\sigma$, such that their contribution to the overall loss becomes dominant relative to the FM term. Consequently, the model progressively reduces its reliance on the pseudo–ground truth and instead focuses on learning more informative latent representations that are better tailored to the target task.

\subsection{Curve reconstruction}
\label{res_disc_2}

\begin{figure}[tb]
    \centering
    \includegraphics[width=0.90\linewidth]{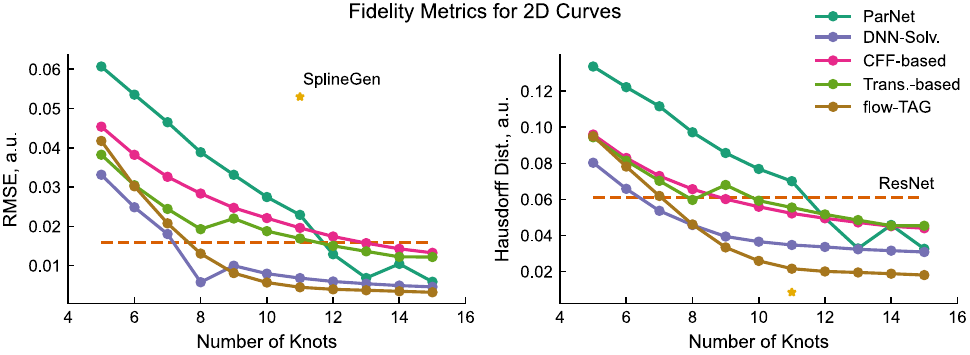}
    \caption{2D curve reconstruction results on the SplineGen dataset; RMSE (left) and Hausdorff distance (right) plotted across different numbers of knots.}
    \label{fig_7}
\end{figure}

We now turn to assessing the suitability of the knot placements produced by our method for B-spline curve interpolation. To this end, we refer to Table~\ref{fig_7}, which reports the results of 2D B-spline interpolation obtained by flow-TAG, alongside several baseline algorithms for comparison.

Our method achieves the lowest RMSE and Haus.~dist. on the test dataset for 9--15 knots. The closest competitor, DNN-Solver, being an optimizing iterative solver, produces better results the longer the optimization runs, and eventually outperforms flow-TAG for 5--8 knots. However, such optimization takes incomparably more time than generating knots with flow-TAG. We report the results for SplineGen with the important caveat that the corresponding performance metrics are taken directly from the original publications. We did not independently reproduce these experiments, as the implementation of SplineGen was not publicly available in open-access form. In terms of Haus.~dist., our method yields to SplineGen at 11 knots. This is an intriguing result, considering that RMSE usually correlates with Haus.~dist., while the RMSE of SplineGen turns out to be an order of magnitude higher than that of our method. 

\begin{figure}[tb]
    \centering
    \includegraphics[width=1.0\linewidth]{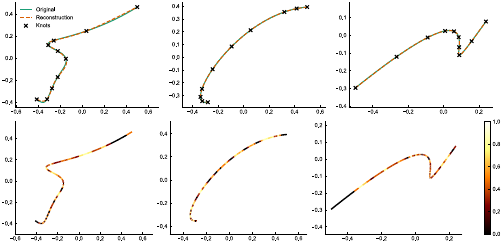}
    \caption{Examples of B-spline 2D curve approximations with knots outputted by flow-TAG (upper row) and uncertainty map for the corresponding curves estimated across 10 generations (lower row).}
    \label{fig_8}
\end{figure}

Figure~\ref{fig_8} shows three examples from the 2D curve test set together with B-spline interpolations based on the knots generated by flow-TAG. One can observe that the model tends to place knots in regions with higher curvature. However, unlike typical heuristic algorithms, it does not always position a knot exactly at the point of maximum curvature. Nevertheless, the interpolation quality does not suffer. The lower row of plots in Figure~\ref{fig_8} displays the uncertainty maps for each curve. The values were calculated as the standard deviation of the soft masks averaged across 10 independent generations. The integrated uncertainty estimation capability constitutes a key strength of the proposed framework, as it provides quantitative insight into the model’s confidence when approximating a target curve. In particular, Figure~\ref{fig_8} illustrates that the model exhibits heterogeneous uncertainty regarding the placement of individual knots, with some knots being associated with substantially higher uncertainty than others. Moreover, distinct regions of low uncertainty are clearly discernible; these regions were systematically avoided during the knot placement process.

\begin{figure}[tb]
    \centering
    \includegraphics[width=0.9\linewidth]{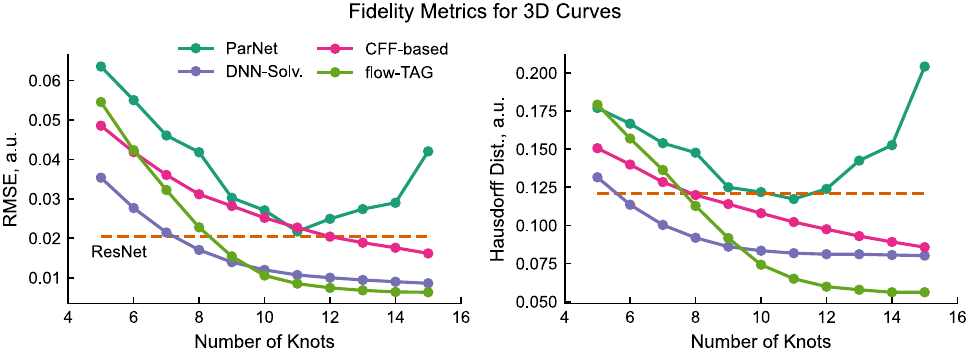}
    \caption{3D curve reconstruction results on the SplineGen dataset; RMSE (left) and Hausdorff distance (right) plotted across different numbers of knots.}
    \label{fig_9}
\end{figure}

\begin{figure}[tb]
    \centering
    \includegraphics[width=1.0\linewidth]{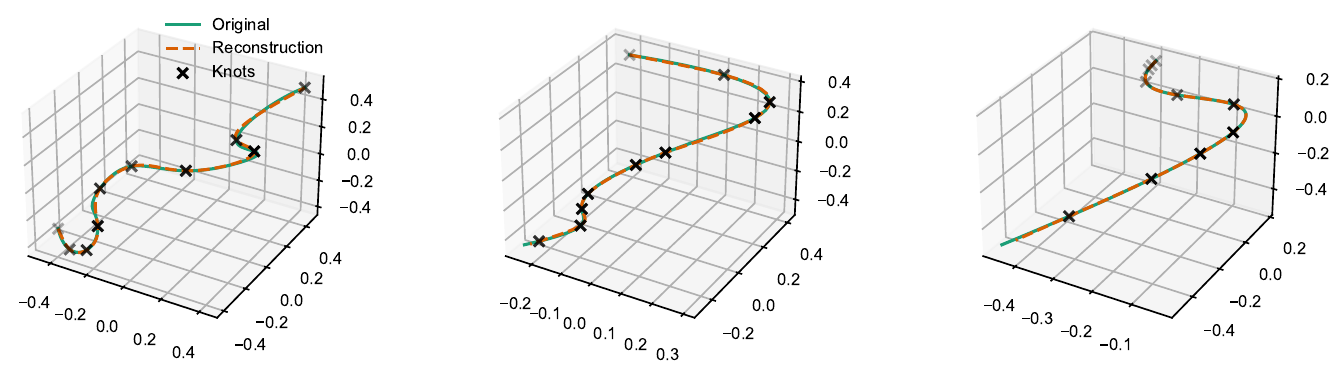}
    \caption{Examples of B-spline 3D curve approximations with knots outputted by flow-TAG.}
    \label{fig_10}
\end{figure}

On the test 3D SplineGen set (see Figure~\ref{fig_9}), our method achieves the best interpolation quality both in terms of RMSE and Haus.~dist. values for 10--15 knots, as compared to the baselines. For fewer knots, flow-TAG outputs suboptimal placements associated with higher error values. We do not report the results for SplineGen evaluation on 3D curves, since they were not provided in the original works. For visual inspection of the interpolation, we present B-spline interpolation of 3 random 3D curves in Figure~\ref{fig_10}.

\subsection{Curve compression}
\label{res_disc_3}

We now examine the direct integration of flow-TAG within an ECG compression framework. To this end, we have trained the flow-TAG components on the training subset of MIT-BIH Arrhythmia database. Next, we developed a simple lossy compression scheme, whose procedural pipeline can be concisely summarized as follows:
\begin{enumerate}
    \item ECG encoding by flow-TAG $\rightarrow$ signal values at knot and knot locations.
    \item Encoding the binary knot placement mask $\rightarrow$ bitstream.
    \item Non-linear quantization of the values at the knots~(Lloyd’s algorithm) followed by the entropy encoding (Huffman codes) $\rightarrow$ bitstream.
\end{enumerate}
Table~\ref{tab:ecg_comp_performance} shows that the proposed setup provides a compression ratio of 13 with acceptable signal distortion below 9\% (the referred result was obtained for 30 internal knots retained per patch and 30 nonlinear quantization levels). When comparing the efficiency with several established baseline methods (see Table~\ref{tab:ecg_comp_performance}), it becomes evident that the proposed approach achieves a moderate level of compression, outperforming, for instance, the method reported in the paper of \cite{mohebbian_ecg_2023}. Analysis of the quality scores indicates that the proposed algorithm outperforms the ECG-adapted JPEG2000, CAE from the study of \cite{yildirim_efficient_2018}, and the recent multichannel CAE of Bekiryazici~et~al.~\cite{bekiryazici_novel_2025}. These results clearly demonstrate the fundamental viability of flow-TAG for ECG compression. Examples of an ECG signal reconstructed after compression by flow-TAG are shown in Figure~\ref{fig_10}. The model allocates knots in the regions corresponding to the characteristic waves, yielding a robust latent representation of the ECG signal.

\begin{table}[t]
\centering
\caption{Comparison of the ECG compression performance of the flow-TAG-based algorithm with different baselines.}
\label{tab:ecg_comp_performance}
\begin{tabular}{lccc}
\toprule
\textbf{Method} & \textbf{PRDN, \%} & \textbf{Comp. ratio} & \textbf{QS} \\
\midrule
Mohebbian~et~al.~\cite{mohebbian_ecg_2023}* & 2.3 & 7.8 & 3.39 \\
Bekiryazici~et~al.~\cite{sharma_ecg_2024} & 8.21 & 33.11 & 4.03 \\
Bekiryazici~et~al.~\cite{bekiryazici_novel_2025} & 9.85 & 20.23 & 2.05 \\
JPEG2000 & 16.76 & 19.48 & 1.16 \\
Yildirim~et~al.~\cite{yildirim_efficient_2018} & 31.17 & 32.25 & 1.03 \\
\midrule
\textbf{Ours} & 5.13 & 13 & 2.55 \\
\bottomrule
\multicolumn{4}{c}{*~-- unnornalized PRD value from the original paper} \\
\end{tabular}
\end{table}

\begin{figure}[tb]
    \centering
    \includegraphics[width=0.9\linewidth]{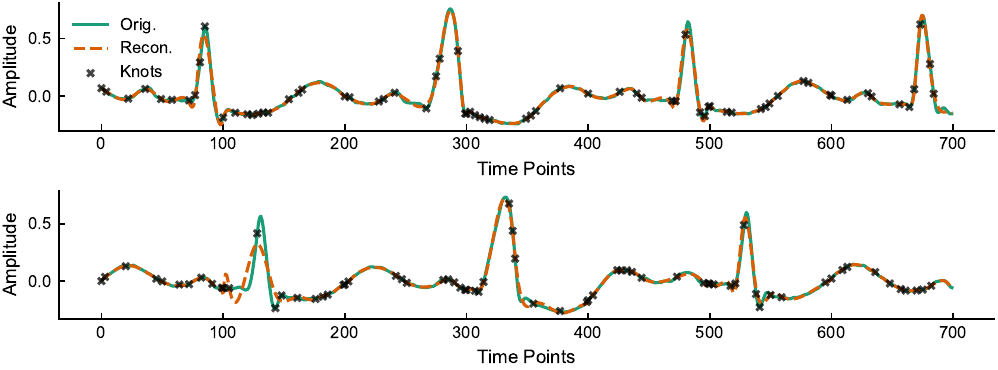}
    \caption{Examples of B-spline ECG approximations with knots outputted by flow-
TAG.}
    \label{fig_12}
\end{figure}

\section{Conclusion}
\label{concl}


This paper proposes flow-TAG framework: a generative flow model based on 1D U-Net for generating the parametrization (knot vectors) for cubic B-spline curve approximation. The framework integrates several novel design components, specifically: 1)~a redefinition of the B-spline functional through a 1D U-Net deep learning architecture; and 2)~a training strategy based on a self-tuning, multi-task, geometry-informed loss function that jointly incorporates the flow matching component and a curve-fidelity component. The resulting generative framework has demonstrated its effectiveness in approximating 2D and 3D curves (public SplineGen dataset) and its advantage over baselines that optimize spline parameterization. In addition, we investigated the prospects of the utilization of flow-TAG as an ECG signal encoder for a compression algorithm. The outcomes of these tests were encouraging and indicate favorable preliminary results.


As an experimental framework, flow-TAG exhibits certain limitations. A primary constraint arises from its patch-wise operation: curves are partitioned into segments of 100 points, for which the corresponding knots are generated. While this strategy can be practical, it may prove suboptimal in specific scenarios. Conversely, the global strategy—processing the entire curve in a single pass—introduces additional constraints on the generative neural network architecture. In principle, however, flow-TAG can be implemented under both~regimes.

Flow-TAG exhibits numerous potential applications in computer-aided design. Among these, we anticipate that particularly promising directions include the generation of latent representations for streaming data and the design of advanced algorithms for ECG signal compression.

\appendix
\section{U-Net design and training details}
\label{app_1}

As mentioned in Section 5.1, the models that constitute flow-TAG framework share a common architecture, summarized in Table~\ref{tab:unet1d_arch}. U-Net depth was set to 4, which is sufficient to cover the receptive field of curves with $N=100$~points. Channel $C=2$~or~$3$ for $\mathcal{D}$ and $f_\theta$, for 2D or 3D curves, respectively; in $v_\phi$, $C=1$. The architecture of the flow model has certain distinctions, such as multiscale features and time conditioning. The summary of it is given in Table~\ref{tab:flow_unet1d_arch}.

\begin{table}
\centering
\caption{1D U-Net architecture used for the denoising autoencoder $\mathcal{D}$ and~task~model~$f_\theta$.}
\label{tab:unet1d_arch}
\setlength{\tabcolsep}{4pt}
\begin{tabular}{llll}
\toprule
\textbf{Level} & \textbf{Main ops} & \textbf{Ch.\,in $\to$ out} & \textbf{Len.} \\
\midrule
Input & - & \(B{\times}C{\times}N\) & \(N\) \\
E1 & DConv{+}Pool$_2$ & \(C \rightarrow 32\) & \(N \rightarrow N/2\) \\
E2 & DConv{+}Pool$_2$ & \(32 \rightarrow 64\) & \(N/2 \rightarrow N/4\) \\
E3 & DConv{+}Pool$_2$ & \(64 \rightarrow 128\) & \(N/4 \rightarrow N/8\) \\
E4 & DConv{+}Pool$_2$ & \(128 \rightarrow 256\) & \(N/8 \rightarrow N/16\) \\
Bottleneck & DConv & \(256 \rightarrow 512\) & \(N/16\) \\
D1 & UpConv{+}Skip{+}DConv & \(512 \rightarrow 256\) & \(N/16 \rightarrow N/8\) \\
D2 & UpConv{+}Skip{+}DConv & \(256 \rightarrow 128\) & \(N/8 \rightarrow N/4\) \\
D3 & UpConv{+}Skip{+}DConv & \(128 \rightarrow 64\) & \(N/4 \rightarrow N/2\) \\
D4 & UpConv{+}Skip{+}DConv & \(64 \rightarrow 32\) & \(N/2 \rightarrow N\) \\
Output & \(1{\times}1\) Conv{+}transpose & \(32 \rightarrow C\) & \(N\) \\
\bottomrule
\end{tabular}

\vspace{2pt}
\footnotesize DConv \(=\) [Conv1D(\(k{=}3,p{=}1\)) + BN + ReLU] \(\times 2\); UpConv \(=\) ConvTranspose1D(\(k{=}2,s{=}2\)); right-padding is used for odd-length skip mismatches.
\end{table}

\begin{table}
\centering
\caption{1D U-Net architecture used for the flow model $v_\phi$.}
\label{tab:flow_unet1d_arch}
\small
\setlength{\tabcolsep}{3pt}
\resizebox{\columnwidth}{!}{%
\begin{tabular}{llll}
\toprule
\textbf{Level} & \textbf{Main ops} & \textbf{Ch.\,in $\to$ out} & \textbf{Len.} \\
\midrule
Input      & —                                          & $B{\times}1{\times}N$          & $N$      \\
TimeEmb    & MLP$(1{\to}d_t)$                           & —                              & $d_t{=}128$ \\
E1         & Cat$(x,t_1)$+DConv+Pool$_2$                & $1{+}64{=}65\to32$             & $N/2$    \\
E2         & Cat$(x,t_2)$+DConv+Pool$_2$                & $32{+}128{=}160\to64$          & $N/4$    \\
E3         & Cat$(x,t_3)$+DConv+Pool$_2$                & $64{+}256{=}320\to128$         & $N/8$    \\
E4         & Cat$(x,t_4)$+DConv+Pool$_2$                & $128{+}512{=}640\to256$        & $N/16$   \\
Bottleneck & Cat$(x,t)$+DConv                           & $256{+}512{=}768\to512$        & $N/16$   \\
D1         & UpConv+Cat$(s,x,t)$+DConv                  & $512{+}512{=}1024\to256$       & $N/8$    \\
D2         & UpConv+Cat$(s,x,t)$+DConv                  & $256{+}256{=}512\to128$        & $N/4$    \\
D3         & UpConv+Cat$(s,x,t)$+DConv                  & $128{+}128{=}256\to64$         & $N/2$    \\
D4         & UpConv+Cat$(s,x,t)$+DConv                  & $64{+}64{=}128\to32$           & $N$      \\
Output     & $1{\times}1$Conv+transp.                   & $32\to1$                       & $N$      \\
\bottomrule
\end{tabular}}
\vspace{2pt}

{\footnotesize $t_i$\,=\,time emb.\ projected to scale $i$; $s$\,=\,skip (cond.\ feat.\ from $\mathcal{D}$ encoder);
DConv\,=\,[Conv1D$(k{=}3,p{=}1)$\,+\,BN\,+\,ReLU]${\times}2$;
UpConv\,=\,ConvTranspose1D$(k{=}2,s{=}2)$.}
\end{table}

$\mathcal{D}$ and $f_\theta$  were trained using the geometry-informed loss~(\ref{eq:eq_27})-(\ref{eq:eq_28}), $v_\phi$ was trained with the multitask learning objective~(\ref{eq:eq_26}) with uncertainty-based loss weighting. The model ingests 2D/3D curves corrupted with additive Gaussian noise $\sigma = 0.01$ and aims to reconstruct the noiseless originals. The training spans 20 epochs with batch ($B=32$) sampling from the SplineGen training set, using the Adam~optimizer~\citep{kingma_adam_2014} with learning rate $\eta=10^{-4}$. Training of flow-TAG framework is summarized in Algorithm~\ref{alg:flow_model_train}.
\begin{algorithm}
\caption{Training of the conditional 1D U-Net flow $v_\phi$ model}
\label{alg:flow_model_train}
\begin{algorithmic}[1]
\REQUIRE Training curves $y \in \mathbb{R}^{B \times N \times C}$, frozen pretrained denoising autoencoder $\mathcal{D}$ and task model $f_\theta$, noise level $\sigma$
\FOR{$\text{iter} = 1$ to \text{epochs}}
    \STATE Sample training batch: $y_B$
    \STATE Add input noise: $y_B^{\mathrm{noisy}} \gets y_B + \sigma \cdot \epsilon, \quad \epsilon \sim \mathcal{N}(0,1)$
    \STATE Get conditioning features: $h_i \gets \mathcal{D}(y_B^{\mathrm{noisy}}), \quad i=\text{depth}+1$
    \STATE Get knot count: $M \sim \mathrm{UniformInt}(5, 15)$ \quad \textcolor{gray}{// single $M$ per batch}
    \STATE Sample $x_0 \sim \mathcal{N}(0, I)^{B \times N \times 1}, \quad t \sim \mathcal{U}(0, 1)^B$
    \STATE Get GT knot mask estimation: $x^{\star} \gets \mathcal{F}(y_B,M)$ \quad \textcolor{gray}{// algorithm of Yeh~et~al.~\cite{yeh_fast_2020}}
    \STATE Define linear FM path and target velocity: $\tilde{x}_t = (1-t)x_0 + t x^{\star},\qquad
    u^{\star} = x^{\star} - x_0$
    \STATE Predict flow velocity and mask logits: $x_t = x_0 + v_\phi(\tilde{x}_t, t, h_i) \cdot t$
    \STATE Evaluate the percentile thresholds: $\tau \gets \mathrm{Quantile}(x_t, 1-M/N)$
    \STATE Get the soft masks: $g(x_t, \tau, t) \gets \mathcal{S} \!\left( s(t) \cdot (x_t - \tau) \right)$
    \STATE Apply soft sampling: $y_B^{\text{soft-masked}} \gets y_B^{\mathrm{noisy}} \odot g$ 
    \STATE Reconstruct with pretrained task model: $\hat{y}_B \gets f_\theta(y_B^{\text{soft-masked}})$
    \STATE Compute the loss and update parameters: $\mathcal{L}(y_B, \hat{y}_B), \quad \nabla_{\phi,\sigma_1,\sigma_2,\sigma_3} \mathcal{L}$
\ENDFOR
\RETURN Trained model $v_\phi$
\end{algorithmic}
\end{algorithm}

\section{Ablation study on the training objective of~$\mathcal{D}$~and~$v_\phi$}
\label{app_2}

In the Table~\ref{tab:denoiser_loss_ablation}~and~\ref{tab:flowtag_loss_ablation} we report the denoising and the flow models' performance after being trained for 10,000 iterations with different combinations of the total training loss, Eq.~(\ref{eq:eq_26}).

\begin{table}
\centering
\caption{Performance of $\mathcal{D}$ trained with the different objectives.}
\label{tab:denoiser_loss_ablation}
\begin{tabular}{lcc}
\toprule
\textbf{Loss component} & \textbf{RMSE, a.u.} & \textbf{Haus. Dist., a.u.} \\
\midrule
$\mathcal{L}_{\text{MSE}}$ & 0.010632 & 0.041932 \\
$\mathcal{L}_{\text{MSE}} + \mathcal{L}_{\text{Smooth}}$ & 0.009503 & 0.037222 \\
$\frac{1}{2\sigma_1^{2}}\mathcal{L}_{\text{MSE}} + \frac{1}{2\sigma_2^{2}}\mathcal{L}_{\text{Smooth}}$ & 0.010928 & 0.034751 \\
\end{tabular}
\end{table}

\begin{table}
\centering
\caption{Performance of $v_\phi$ trained with the different objectives.}
\label{tab:flowtag_loss_ablation}
\begin{tabular}{lcc}
\toprule
\textbf{Loss component} & \textbf{RMSE, a.u.} & \textbf{Haus. Dist., a.u.} \\
\midrule
$\mathcal{L}_{\text{FM}}$ & 0.006007 & 0.025700 \\
$\frac{1}{2\sigma_1^{2}}\mathcal{L}_{\text{MSE}} + \frac{1}{2\sigma_2^{2}}\mathcal{L}_{\text{Smooth}}$ & 0.011002 & 0.039807 \\
$\mathcal{L}_{\text{total}}$ & 0.005056 & 0.022596 \\
\end{tabular}
\end{table}

\bibliographystyle{unsrt}  
\bibliography{references} 

@article{pal_electrocardiogram_2023,
	title = {Electrocardiogram signal compression using adaptive tunable-{Q} wavelet transform and modified dead-zone quantizer},
	volume = {142},
	issn = {0019-0578},
	journal = {ISA transactions},
	publisher = {Elsevier},
	author = {Pal, Hardev Singh and Kumar, A and Vishwakarma, Amit and Lee, Heung-No},
	year = {2023},
	pages = {335--346},
}

@article{singhai_ecg_2023,
	title = {{ECG} signal compression based on optimization of wavelet parameters and threshold levels using evolutionary techniques},
	volume = {42},
	issn = {0278-081X},
	number = {6},
	journal = {Circuits, Systems, and Signal Processing},
	publisher = {Springer},
	author = {Singhai, Paridhi and Kumar, Anil and Ateek, A and Ansari, Irshad Ahmad and Singh, Girish Kumar and Lee, Heung No},
	year = {2023},
	pages = {3509--3537},
}

@article{xu_ecg_2017,
	title = {{ECG} signal de-noising and baseline wander correction based on {CEEMDAN} and wavelet threshold},
	volume = {17},
	issn = {1424-8220},
	number = {12},
	journal = {Sensors},
	publisher = {MDPI},
	author = {Xu, Yang and Luo, Mingzhang and Li, Tao and Song, Gangbing},
	year = {2017},
	pages = {2754},
}

@article{chang_arrhythmia_2010,
	title = {Arrhythmia {ECG} noise reduction by ensemble empirical mode decomposition},
	volume = {10},
	issn = {1424-8220},
	number = {6},
	journal = {Sensors},
	publisher = {Molecular Diversity Preservation International (MDPI)},
	author = {Chang, Kang-Ming},
	year = {2010},
	pages = {6063--6080},
}

@article{lecun_deep_2015,
	title = {Deep learning},
	volume = {521},
	issn = {0028-0836},
	number = {7553},
	journal = {nature},
	publisher = {Nature Publishing Group UK London},
	author = {LeCun, Yann and Bengio, Yoshua and Hinton, Geoffrey},
	year = {2015},
	pages = {436--444},
}

@inproceedings{song_hybrid_2024,
	title = {A {Hybrid} {Parametrization} {Method} for {B}‐{Spline} {Curve} {Interpolation} via {Supervised} {Learning}},
	volume = {43},
	isbn = {0167-7055},
	number = {7},
	publisher = {Wiley Online Library},
	author = {Song, Tianyu and Shen, Tong and Ge, Linlin and Feng, Jieqing},
	year = {2024},
	pages = {e15240},
}

@article{laube_learnt_2018,
	title = {Learnt knot placement in {B}-spline curve approximation using support vector machines},
	volume = {62},
	issn = {0167-8396},
	journal = {Computer Aided Geometric Design},
	publisher = {Elsevier},
	author = {Laube, Pascal and Franz, Matthias O and Umlauf, Georg},
	year = {2018},
	pages = {104--116},
}

@article{hughes_isogeometric_2005,
	title = {Isogeometric analysis: {CAD}, finite elements, {NURBS}, exact geometry and mesh refinement},
	volume = {194},
	issn = {0045-7825},
	number = {39-41},
	journal = {Computer methods in applied mechanics and engineering},
	publisher = {Elsevier},
	author = {Hughes, Thomas JR and Cottrell, John A and Bazilevs, Yuri},
	year = {2005},
	pages = {4135--4195},
}

@article{lasemi_recent_2010,
	title = {Recent development in {CNC} machining of freeform surfaces: {A} state-of-the-art review},
	volume = {42},
	issn = {0010-4485},
	number = {7},
	journal = {Computer-Aided Design},
	publisher = {Elsevier},
	author = {Lasemi, Ali and Xue, Deyi and Gu, Peihua},
	year = {2010},
	pages = {641--654},
}

@article{langeron_new_2004,
	title = {A new format for 5-axis tool path computation, using {Bspline} curves},
	volume = {36},
	issn = {0010-4485},
	number = {12},
	journal = {Computer-Aided Design},
	publisher = {Elsevier},
	author = {Langeron, Jean Marie and Duc, Emmanuel and Lartigue, Claire and Bourdet, Pierre},
	year = {2004},
	pages = {1219--1229},
}

@book{prautzsch_bezier_2002,
	title = {Bézier and {B}-spline techniques},
	volume = {6},
	publisher = {Springer},
	author = {Prautzsch, Hartmut and Boehm, Wolfgang and Paluszny, Marco},
	year = {2002},
}

@article{pottmann_industrial_2005,
	title = {Industrial geometry: recent advances and applications in {CAD}},
	volume = {37},
	issn = {0010-4485},
	number = {7},
	journal = {Computer-Aided Design},
	publisher = {Elsevier},
	author = {Pottmann, Helmut and Leopoldseder, Stefan and Hofer, Michael and Steiner, Tibor and Wang, Wenping},
	year = {2005},
	pages = {751--766},
}

@article{bilgin_compression_2003,
	title = {Compression of electrocardiogram signals using {JPEG} 2000},
	volume = {49},
	issn = {0098-3063},
	number = {4},
	journal = {IEEE Transactions on Consumer Electronics},
	author = {Bilgin, Ali and Marcellin, Michael W and Altbach, Maria I},
	year = {2003},
	pages = {833--840},
}

@article{scholz_parameterization_2021,
	title = {Parameterization for polynomial curve approximation via residual deep neural networks},
	volume = {85},
	issn = {0167-8396},
	journal = {Computer Aided Geometric Design},
	publisher = {Elsevier},
	author = {Scholz, Felix and Juettler, Bert},
	year = {2021},
	pages = {101977},
}

@article{yildirim_efficient_2018,
	title = {An efficient compression of {ECG} signals using deep convolutional autoencoders},
	volume = {52},
	issn = {1389-0417},
	journal = {Cognitive Systems Research},
	publisher = {Elsevier},
	author = {Yildirim, Ozal and San Tan, Ru and Acharya, U Rajendra},
	year = {2018},
	pages = {198--211},
}

@article{sharma_ecg_2024,
	title = {{ECG} compression based on empirical mode decomposition and tunable-{Q} wavelet transform with validation using heartbeat classification},
	volume = {18},
	issn = {1863-1703},
	number = {4},
	journal = {Signal, Image and Video Processing},
	publisher = {Springer},
	author = {Sharma, Neenu and Sunkaria, Ramesh Kumar},
	year = {2024},
	pages = {3079--3095},
}

@article{bekiryazici_novel_2025,
	title = {A novel multichannel sparse convolutional autoencoder for electrocardiogram signal compression},
	issn = {0022-0736},
	journal = {Journal of Electrocardiology},
	publisher = {Elsevier},
	author = {Bekiryazıcı, Tahir and Damkacı, Mehmet and Aydemir, Gürkan and Gürkan, Hakan},
	year = {2025},
	pages = {154125},
}

@article{zigel_weighted_2002,
	title = {The weighted diagnostic distortion ({WDD}) measure for {ECG} signal compression},
	volume = {47},
	issn = {0018-9294},
	number = {11},
	journal = {IEEE transactions on biomedical engineering},
	publisher = {IEEE},
	author = {Zigel, Yaniv and Cohen, Arnon and Katz, Amos},
	year = {2002},
	pages = {1422--1430},
}

@article{moody_impact_2001,
	title = {The impact of the {MIT}-{BIH} arrhythmia database},
	volume = {20},
	issn = {0739-5175},
	number = {3},
	journal = {IEEE engineering in medicine and biology magazine},
	publisher = {IEEE},
	author = {Moody, George B and Mark, Roger G},
	year = {2001},
	pages = {45--50},
}

@inproceedings{kendall_multi-task_2018,
	title = {Multi-task learning using uncertainty to weigh losses for scene geometry and semantics},
	author = {Kendall, Alex and Gal, Yarin and Cipolla, Roberto},
	year = {2018},
	pages = {7482--7491},
}

@article{williams_unified_2023,
	title = {A unified framework for {U}-{Net} design and analysis},
	volume = {36},
	journal = {Advances in Neural Information Processing Systems},
	author = {Williams, Christopher and Falck, Fabian and Deligiannidis, George and Holmes, Chris C and Doucet, Arnaud and Syed, Saifuddin},
	year = {2023},
	pages = {27745--27782},
}

@book{prenter_splines_2008,
	title = {Splines and variational methods},
	isbn = {0-486-46902-6},
	publisher = {Courier Corporation},
	author = {Prenter, Paddy M},
	year = {2008},
}

@article{unser_splines_2002,
	title = {Splines: {A} perfect fit for signal and image processing},
	volume = {16},
	issn = {1053-5888},
	number = {6},
	journal = {IEEE Signal processing magazine},
	publisher = {IEEE},
	author = {Unser, Michael},
	year = {2002},
	pages = {22--38},
}

@book{de_boor_practical_1978,
	title = {A practical guide to splines},
	volume = {27},
	isbn = {0-387-90356-9},
	publisher = {Springer New York},
	author = {De Boor, Carl},
	year = {1978},
}

@article{zhu_differentiable_2025,
	title = {Differentiable {Fast} {Top}-{K} {Selection} for {Large}-{Scale} {Recommendation}},
	journal = {arXiv preprint arXiv:2510.11472},
	author = {Zhu, Yanjie and Zhang, Zhen and Wang, Yunli and Wang, Zhiqiang and Li, Yu and Zhou, Rufan and Wen, Shiyang and Jiang, Peng and Lin, Chenhao and Yang, Jian},
	year = {2025},
}

@inproceedings{pavelkin_spline-based_2025,
	title = {Spline-{Based} {Shape} {Compression} for {Interventional} {Device} {Tracking}},
	publisher = {Springer},
	author = {Pavelkin, Roman and Zavala-Mondragon, Luis A and Ekin, Ahmet and Sommen, Fons van der},
	year = {2025},
	pages = {179--192},
}

@article{kingma_adam_2014,
	title = {Adam: {A} method for stochastic optimization},
	journal = {arXiv preprint arXiv:1412.6980},
	author = {Kingma, Diederik P},
	year = {2014},
}

@inproceedings{ronneberger_u-net_2015,
	title = {U-net: {Convolutional} networks for biomedical image segmentation},
	publisher = {Springer},
	author = {Ronneberger, Olaf and Fischer, Philipp and Brox, Thomas},
	year = {2015},
	pages = {234--241},
}

@book{lai_principles_2025,
	title = {The {Principles} of {Diffusion} {Models}},
	author = {Lai, Chieh-Hsin and Song, Yang and Kim, Dongjun and Mitsufuji, Yuki and Ermon, Stefano},
	year = {2025},
}

@article{lipman_flow_2022,
	title = {Flow matching for generative modeling},
	journal = {arXiv preprint arXiv:2210.02747},
	author = {Lipman, Yaron and Chen, Ricky TQ and Ben-Hamu, Heli and Nickel, Maximilian and Le, Matt},
	year = {2022},
}

@book{piegl_nurbs_2012,
	title = {The {NURBS} book},
	isbn = {3-642-59223-6},
	publisher = {Springer Science \& Business Media},
	author = {Piegl, Les and Tiller, Wayne},
	year = {2012},
}

@article{saillot_b-spline_2024,
	title = {B-spline curve approximation with transformer neural networks},
	volume = {223},
	issn = {0378-4754},
	journal = {Mathematics and Computers in Simulation},
	publisher = {Elsevier},
	author = {Saillot, Mathis and Michel, Dominique and Zidna, Ahmed},
	year = {2024},
	pages = {275--287},
}

@article{wen_deep_2024,
	title = {The deep neural network solver for {B}-spline approximation},
	volume = {169},
	issn = {0010-4485},
	journal = {Computer-Aided Design},
	publisher = {Elsevier},
	author = {Wen, Zepeng and Luo, Jiaqi and Kang, Hongmei},
	year = {2024},
	pages = {103668},
}

@inproceedings{laube_deep_2018,
	title = {Deep learning parametrization for {B}-spline curve approximation},
	isbn = {1-5386-8425-X},
	publisher = {IEEE},
	author = {Laube, Pascal and Franz, Matthias O and Umlauf, Georg},
	year = {2018},
	pages = {691--699},
}

@article{yeh_fast_2020,
	title = {Fast automatic knot placement method for accurate {B}-spline curve fitting},
	volume = {128},
	issn = {0010-4485},
	journal = {Computer-aided design},
	publisher = {Elsevier},
	author = {Yeh, Raine and Nashed, Youssef SG and Peterka, Tom and Tricoche, Xavier},
	year = {2020},
	pages = {102905},
}

@article{zou_splinegen_2025,
	title = {{SplineGen}: {Approximating} unorganized points through generative {AI}},
	volume = {178},
	issn = {0010-4485},
	journal = {Computer-Aided Design},
	publisher = {Elsevier},
	author = {Zou, Qiang and Zhu, Lizhen and Wu, Jiayu and Yang, Zhijie},
	year = {2025},
	pages = {103809},
}

@article{mohebbian_ecg_2023,
	title = {{ECG} compression using optimized {B}-spline},
	volume = {82},
	issn = {1380-7501},
	number = {14},
	journal = {Multimedia Tools and Applications},
	publisher = {Springer},
	author = {Mohebbian, Mohammad Reza and Wahid, Khan A},
	year = {2023},
	pages = {21071--21083},
}

@book{farin_curves_2001,
	title = {Curves and surfaces for {CAGD}: a practical guide},
	isbn = {0-08-050354-3},
	publisher = {Elsevier},
	author = {Farin, Gerald},
	year = {2001},
}

\end{document}